\documentclass[11pt,twoside]{article}
\usepackage{asp2010}

\resetcounters

\begin{document}

\title{The Formation of Close Binaries}
\author{Kaitlin M. Kratter$^1$
\affil{$^1$Harvard-Smithsonian Center for Astrophysics, 60 Garden Street, Cambridge, MA 02138, USA}}

\begin{abstract}
Binary stars produce an array of dramatic astrophysical phenomena. They allow us to probe stellar structure, nuclear physics, and gravitational wave physics. They also produce the powerful supernovae that allow us to measure the scale of the universe. Despite their importance and ubiquity, many questions remain unanswered as to how the star formation process produces the wide array of stellar multiples that we observe. A complete model for binary formation encompasses three main components. We must know: (1) the primordial population of systems, (2) the influence of the dynamical processes that reshape this distribution as stars form and natal clusters disperse, and (3) the role of binary stellar evolution. In this article I review the most prominent theories for binary formation: turbulent core fragmentation, disk fragmentation, and competitive accretion. I argue that turbulent core fragmentation at all masses, with disk fragmentation added in at the upper end of the mass spectrum, might explain the trend towards increasing multiplicity for higher mass stars. In addition, I provide a brief overview of the observational statistics and of some of the important processes that modify the primordial distribution of stellar orbits.
\end{abstract}

\section {Introduction}
The study of binarity dates back at least 250 years to \cite{Mitchell1767}, who pointed out the abundance of double stars. Many tout the 21st century as the era of precision cosmology, but we are also entering the era of precision in more classic astronomy -- understanding the ordering of the stars. Great technological advances including  the use of coronography \citep{Hinkley:2007} and adaptive optics \citep{Lafreniere:2009} have brought stellar companions into better focus.

Theories of star formation have been influenced by our heliocentric bias. Until recently, our models have reflected that we live next to a relatively low mass star (apparently) without a stellar or substellar companion. The seminal work of \cite{Duquennoy:1991} on the multiplicity of sun-like stars showed that our sun is somewhat unusual in being alone. Indeed if we consider planets as companions, very few stars are truly isolated \citep{Youdin:2011}. Recent work by \cite{Raghavan2010} has confirmed that roughly half of sun-like stars are in binary or higher order multiple systems. Although the statistics are incomplete for stars of other masses, it seems clear that there is a strong correlation between stellar mass and multiplicity \citep{Lada06, {Mason:2009}}.  The M stars, the most common stars in the galaxy, are usually in single systems. As primary mass increases to just a few solar masses, binaries become ubiquitous, even as the stars themselves become quite rare. The data are consistent with all O and B stars having at least one stellar companion at birth. The statistics on higher order multiplicity are still incomplete. It remains difficult to detect relatively close, extreme mass ratio pairs, particularly for massive primaries. Upcoming searches with next generation telescopes such as ALMA and JWST may improve these statistics.

Armed with this new data, it is time to reconsider our models of single and multiple star formation. To be sure, conclusions drawn from the current data require many caveats. We do not have a complete picture of companions at all separations and mass ratios.  Moreover, given the diverse nature of these systems, it is unlikely that one mechanism can explain them all. This review aims to provide a broad, though necessarily incomplete, picture of the state of the field.

I begin by reviewing theories for binary formation, and proceed to discuss the current observational constraints on both the primordial and field populations. Tightening our focus to systems that might serve as progenitors of compact binaries requires the consideration of another set of processes: those that alter the orbital configurations of the birth population. In this review I will touch only on mechanisms that operate during the main phase of star formation, leaving late stage evolution to other chapters in the proceedings.

A detailed examination of the properties of observed binaries is beyond the scope of this article. For an update to date observational census of low and high mass multiple systems see \cite{Raghavan2010} (low mass) and \cite{Sana:2010} (high mass). For a discussion of higher order multiples, see also \cite{Tokovinin:2008}.

\section{Theories of Binary and Multiple Star Formation}
There are several prominent theories of binary and multiple star formation in the literature. The formation mechanisms can be divided roughly into three categories (1) core mediated processes, (2) disk mediated processes, and (3) few-body dynamical processes. In this review, I neglect channels of binary formation that occur after star formation has ended. There is strong observational evidence that most binary formation  is concurrent with star formation: the binarity fraction for pre-main sequence stars is higher than that for main sequence stars \citep{Mathieu:1994,Kraus:2011}.

\subsection{ Core Mediated Theories}
 The most common mechanisms in the literature fall into the first category, and are described alternately as prompt or early fragmentation, or more specifically as core or turbulent fragmentation \citep{Hoyle:1953,Tsuribe:1999a,Boss:2000,Padoan:2002,{Fisher:2002}}.   The earliest fragmentation scenario is that of \cite{Hoyle:1953} who argued for opacity-limited hierarchical fragmentation, whereby an unstable cloud keeps fragmenting at the local Jeans mass, which decreases as the cloud collapses to  higher densities, until the smallest fragments become optically thick and can no longer cool efficiently. Although this mechanism is inconsistent with the observed stellar mass function (for a thorough review see \citealt{Fisher:2002}), it gave birth to more modern theories of core fragmentation. Another classic binary formation theory is known as core fission, where a core deforms through a series of equilibrium ellipsoidal figures. Cores contract on too short a timescale (compared to their viscous timescale) to evolve in this manner (see  also \cite{Tohline_binrev}.
  
 In the subsequent generation of ``core" fragmentation scenarios, a collapsing, bound gas clump was thought to fragment into two or more objects depending on the ratio of thermal to gravitational energy, $\alpha = 5 c_s^2R/GM$, and the ratio of rotational to gravitational energy, $\beta = \Omega^2 R^3/3GM$m where $c_s$ is the sound speed, $R$ is the core radius, $M$ the core mass, and $\Omega$ the orbital frequency of the core.  Inherent in this parameterization is the assumption that any dynamically important magnetic fields have already diffused out (or can be included simply as an extra effective pressure in the calculation of $\alpha$). When the product $\alpha \beta <0.12$, cores were initially thought unstable \citep{Inutsuka:1992}, although this model has the somewhat counterintuitive result that slowly rotating (small $\beta$) virialized cores remain prone to fragmentation. More recent work has shown this criterion to be incomplete, requiring also that $\alpha < 0.5$ \citep{Tsuribe:1999}. Yet this model neglects many thermal effects as the gas collapses to higher densities.%, and would seem to over produce short period binaries.
 
  The source of angular momentum setting $\beta$ in the model above was thought classically to be galactic shear (e.g. \citealt{Bodenheimer:1995}, and references therein), but the ubiquitous turbulence in the interstellar medium could also be responsible for the observed line-width size relations interpreted as evidence of core rotation \citep{Fisher04,{Goodman:1993}}. 
  
  This identification between rotation and turbulence has led to yet another class of models, appropriately termed turbulent fragmentation. The turbulence scenario suggests that non-linear perturbations expected in a turbulent cloud can cause a sub-region within a core to become over dense and collapse more rapidly than the free-fall timescale of the background core, thereby leading to the production of a secondary condensation within the bound core. Alternatively, turbulent motions can lead to the formation of filamentary structures, which then fragment into multiple objects. These stars are presumed to accrete from their natal core mostly independently. Galactic shear may help stir the turbulence in molecular clouds, although stellar feedback can also be important \citep{Fall:2010}.
  
  Recent work by \cite{Offner:2009} and \cite{OKMKK10} have demonstrated that the turbulent fragmentation mechanism is the dominant source of binaries in radiation hydrodynamic simulations of low mass ($M < M_\odot$) star formation.  These simulations also show rapid orbital migration over the first few tens of thousands of years following formation. Although the binaries were born on scales of  $\sim 0.1$pc, their separations had shrunk to less than $1000$ AU after $\sim 10^4$ years. The final separation distribution produced by turbulent fragmentation remains unknown: simulations which follow the interaction of the gas with stars and the radiation field remain too computationally expensive to run for millions of years. However, one expects the mass ratio distribution to mirror the IMF if the cores accrete onto their own stars relatively unmolested by the binary partner. 
   
 Note that there is another class of star formation models dubbed turbulent fragmentation or ``gravoturbulent fragmentation" which refers to the fragmentation of a molecular cloud on larger scales into many bound objects which form their own stellar systems \citep{Ballesteros-Paredes:2007}. The turbulence sets the mass-scale of the cores, and their turbulent structures, but for the purposes of multiple formation we are concerned with the subsequent fragmentation of the turbulent clump, not the process that generated it.
  
\subsection{Disk Driven Formation}
 Models for binary and multiple formation at a slightly later evolutionary state involve protostellar disks. In the disk fragmentation  model \citep{ARS89,Laughlin:1994,Bonnell:1994,Bonnell:1994b} stars form sufficiently massive protostellar disks that become susceptible to gravitational instabilities. If disks become unstable, and the gas can cool efficiently, they can fragment to produce one or more coplanar companions that accrete from the parent disk, and depending on the formation epoch, the primary's natal cloud. 
 
 Disk fragmentation has received much attention from numericists in recent years.  Prior work on this topic has been severely limited by two assumptions. The first is that disks can be modeled in isolation. Disks modeled in the absence of ongoing accretion are much less likely to become unstable. Ongoing accretion at the outer edge of the disk is nearly always necessary to drive instability \citep{KMK08,KMCY10,KMC11}.  

The second related assumption was that disks were always low in mass relative to the central star. Certainly observed T-Tauri disks tend to be lower in mass, but as observations improve we are beginning to observe more massive disks at early epochs of formation \citep{Andrews:2007,Andrews:2009}. Together these two notions led most previous authors to conclude that disk fragmentation might occur rarely, and could only produce very unequal mass ratio objects, which was inconsistent with binary statistics. 

However, if disks are more massive and undergo continued accretion, simulations suggest that fragmentation may be more common, and that secondaries born in the disk can accrete quickly to catch up to the primary in mass (\citealt{Bonnell:1994a,KMKK10}, though see \citealt{Hanawa:2010}). 
Together with observations of more massive disks, this suggests that the disk fragmentation scenario can in fact produce a wide range of mass ratios. We predict that disk fragmentation is most relevant for systems with primaries more massive than our sun, however recent observations of young clusters indicate disk fragmentation might be relevant at lower masses \citep{Kraus:2011}. In particular, \citet{KMKK10} showed that two dimensionless parameters, which measure the infall rate compared to the sound speed ($\xi = \dot{M}/(c_s^3/G)$, hereafter normalized accretion rate) and the infall rate compared to the orbital time ($\Gamma = \dot{M}/(M_{*d} \Omega$, hereafter normalized rotation period), can accurately predict when disk fragmentation occurs. We show a compilation of these results in Figure 1.

Disk fragmentation is a promising mechanism for the formation of relatively close binaries, because the separations at birth are less than $1000$AU, and often within a hundred AU. Due to interactions with their natal circumstellar or circumbinary disks (see \S\ref{sec-dyn}), significant migration can also occur, much as in the case of planets \citep{Goldreich:1979,Artymowicz:1996}. 

 Disk fragmentation has also been shown to produce higher order, hierarchical multiples \citep{Krumholz2007a,2009SciKrum,Stamatellos:2011}. Such systems may be ideal progenitors to compact binaries (see Perets \& Kratter 2011, submitted).
 
 As with the turbulent fragmentation model, disk fragmentation has yet to provide strong predictions for the separation distribution of
binaries (or multiples). In addition, the mass ratio distribution is unclear. While a range of outcomes are possible, the mass ratios will be regulated by the relative timescale of fragmentation with respect to the accretion time from the core. In systems where fragmentation happens early, stars are likely to approach equal mass as they share infalling material; in contrast, systems that only fragment at the tail end of the star formation process will likely remain at disparate masses. The latter case might lead to the production of brown-dwarf companions \citep{Stamatellos:2009a}.
 
 \begin{figure}
\includegraphics[scale=0.5]{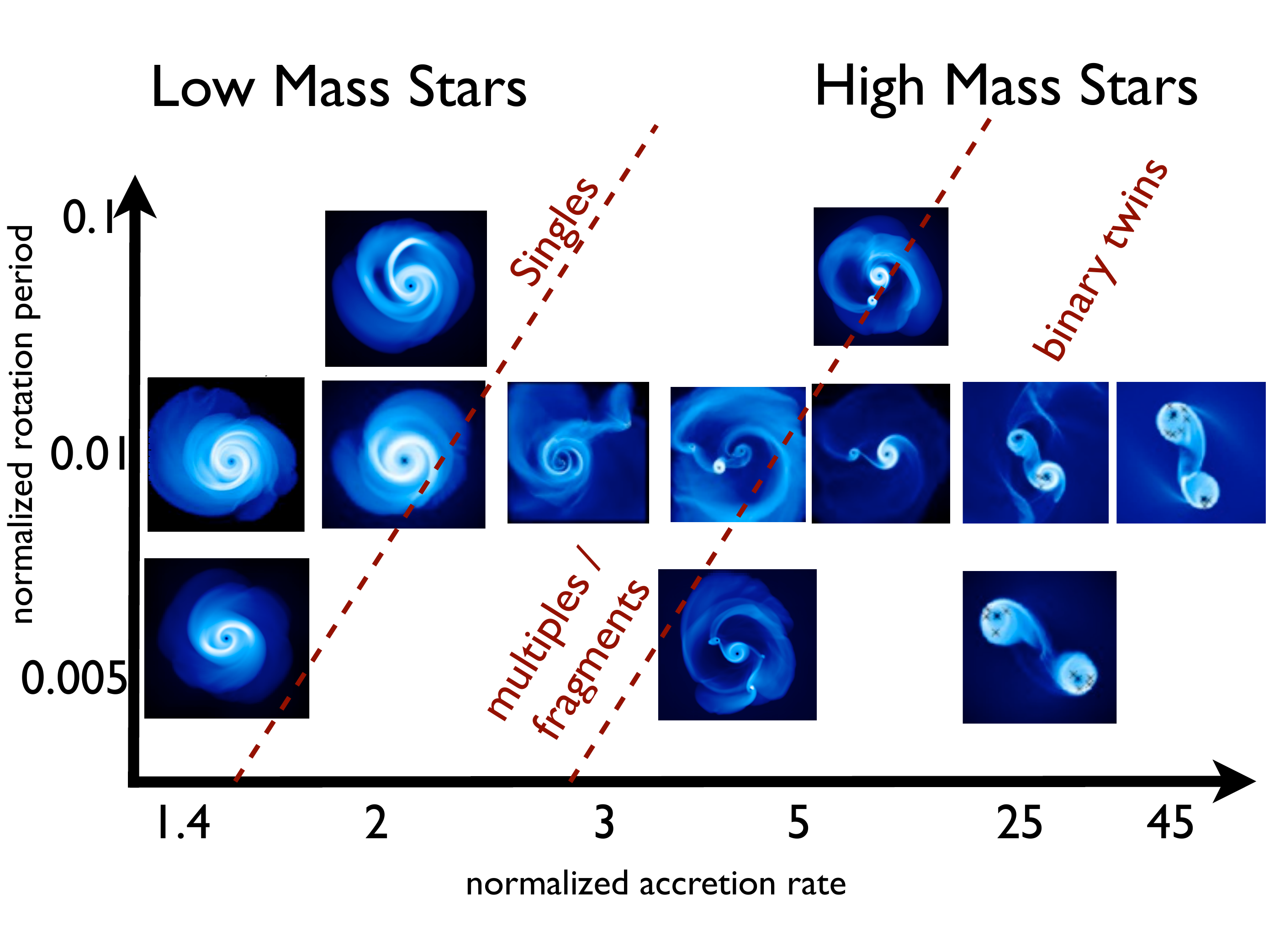}

\caption{Illustration of different disk outcomes as a function of dimensionless accretion rates defined in \cite{KMKK10}. The disks on the right hand side have infall properties characteristic of more massive stars ($M > 2M_\odot$), while those on the left hand side better resemble low mass stars. The increasing propensity towards disk fragmentation at higher masses might explain the increased multiplicity fraction for stars of types earlier than A.}
\end{figure}

 \subsection{Multiple Formation in a dynamically evolving cluster}
 A final  mechanism for multiple formation is that put forth in a series of papers including \cite{Bate:2000}, \cite{Bate:2003}, \cite{Bon2004}, and \cite{Clark:2008a}. These modelers suggest that star clusters form through a dynamic process dubbed competitive accretion. In this scenario, a turbulent molecular cloud forms many roughly Jeans mass clumps that interact with one another and compete for mass from the background cloud. In this environment there are multiple mechanisms for binary formation, including the disk methods mentioned above, as well as dynamical interactions between three or more bodies which we discuss below. In addition, dynamical friction between the protostars and background gas is thought to tighten wider binaries.  What distinguishes this scenario from those listed above is that a stellar system does not accrete from a fixed, marginally bound core, but rather migrates around in a cloud collecting previously unbound material. 
 
 A model put forth by \cite{Goodwin:2005} relies on a combination of the core fragmentation and dynamical evolution hypothesis to reproduce the high multiplicity fraction of young stars in clusters, compared to their field star counterparts. Although they do not specify the precise mechanism for fragmentation, these authors imagine that most cores either break up into 2 or $3+$ individual stars. Through dynamical interaction, the higher order multiples evolve into  single star systems, tight binaries, and stable hierarchical systems. They posit that different cluster environments (which have different rates of close stellar encounters) will naturally evolve to show different multiplicity fractions.
 
 A related mechanism is disk capture.  In the capture scenario, a massive disk increases the gravitational cross-section for capturing a nearby star into a bound orbit. Although initial investigations found this process to be highly inefficient around low mass stars at reasonable stellar densities \citep{Clarke:1991}, more recent work suggests that the process might be better suited to massive stars with correspondingly more massive disks \citep{Moeckel:2007}. However, because this process remains inefficient even in dense clustered environments, it is unlikely to be the dominant formation channel.

With these models in hand, we now turn to mechanisms that reshape the primordial distributions. 

\section{From Birth to Death: Mechanisms for Evolution}\label{sec-dyn}
Two types of interaction are likely to modify the birth population of binaries and multiples while the star formation process is ongoing: (1) accretion disk interactions, and (2) dynamical interactions in few body systems. We address the role of each of these processes in making compact binary progenitors. As noted above, molecular gas may also play a role in shrinking the separations of newborn binaries, however this could well be considered part of the core fragmentation process rather than a separate mechanism since this process operates before the stars have well defined orbital parameters.

\subsection{Disk Interactions}
Binaries born via any mechanism, and especially disk fragmentation can reside within a circumbinary disk of material as they accrete from the background cloud. Gravitational torques can expel the circumbinary disk carrying away a large fraction of the orbital  angular momentum \citep{{Syer:1995},Artymowicz:1996}. The extent to which this process shrinks binary orbits is uncertain, and depends on the mass ratio between the stars and the disk. Some simulations suggest that once the disk is expelled to roughly twice the orbital separation of the binary, orbital evolution stalls, leaving  a stable circular binary \citep{Artymowicz:1996}. This mechanism can likely bring binaries down to separations of 10's of AU, and possible much closer. 

Sometimes, both stars in a binary retain their own circumstellar disks. In this case, tidal interactions between each of the two circumstellar disks allows for transfer of angular momentum from the disks to the orbit, expanding it as in the case of the Earth-Moon system where the rotation of the disk / Earth is faster than that of the orbit of the binary star / Moon. The latter interaction becomes important only when the protostellar disk fills the effective Roche radius of the star-disk system, and so may not be the dominant driver of orbital evolution. Orbital expansion may also depend on the eccentricity of this initial orbit. Note that binaries born from disk fragmentation may well be initially eccentric due to the m=1 mode perturbations that can seed the disk \citep{Sling1990,Krumholz2007a,KMKK10}.
 
 \subsection{Dynamical Interactions in Few Body Systems}
 As mentioned above in the context of the competitive accretion models, dynamical interactions in small-n systems may be extremely important in reshaping the orbital distributions of young systems.  One dominant dynamical mechanism for producing close binaries may be interactions among three or more stars. Binary-binary and single-binary encounters can lead to the ejection of  (usually) the lowest mass star, leaving behind a tighter pair. Similarly, triple and higher multiplicity systems may form or evolve into unstable hierarchical configurations, where their relative orbits interact and trigger chaotic orbital evolution \citep{Mardling:2001}.  Chaotic orbits can lead to ejections, tightening of the remaining binary, partner exchanges, eccentricity excitation, and in some cases, stellar collisions. These dynamical interactions are often seen in the competitive accretion simulations, and reproduce some features of the observed binary distribution of very low mass objects \citep{Dupuy:2011}. 
 
An even broader range of dynamical interactions may occur throughout the lifetime of the binary, particular in dense stellar clusters.  For a more detailed analysis of the different processes at work, and population synthesis calculations of interaction rates, see for example \cite{Portegies-Zwart:1997}, \cite{Pooley:2003},  and \cite{Fregeau:2004}

We now explore the observational evidence for these theories in both the pre-main sequence and field populations.

\section{Linking Theory with Observations }
Advances in numerical techniques mean that each of the formation and evolutionary processes described above has been simulated at high resolution (see references above). Though numerical artifacts remain, many of the mechanisms cannot be ruled out on basic physical grounds. Instead they must be evaluated based on their ability to reproduce the observables of the star formation process.

An unequivocal determination of the binary formation process remains elusive; it would require that we know the conditions in a given cloud and core over its entire formation timescale. We have made great strides in characterizing the different phases of core formation \citep{Enoch09}, but we do not yet see binary formation in progress \citep{Schnee:2010}. 

We must, however, exercise some caution when assessing theories of binary formation that reproduce observed trends without accounting for the underlying physics. While the effort to create models that reproduce observations is important, we should not neglect detailed modeling of molecular clouds and prestellar cores from which binaries form. 

\subsection{Observed Populations of Binaries and Multiples}
The field population of binary stars has a separation distribution that is log-normal for solar type stars, and bimodal and log-flat $(f(\rm{log}(P) \propto \rm{const}$) for more massive systems \citep{Raghavan2010,Sana:2010,Mason:2009}. For the former population, the separation distribution is log normal, with a peak at $\mu = 5.03$ days, and a standard deviation of  $\sigma_{\rm log P} = 2.28$. 
This leaves a sizable population within $\sim10$ AU that may be brought even closer through binary stellar evolution processes \citep{Iben:1985}.  The high mass stars include a large population of short period systems: as many as  60\% of the systems have orbital periods less than 10 days \citep{Sana:2010}. This overabundance of short period systems may be influenced by selection effects, since it is difficult to find lower mass companions to massive stars at distance of 10's-100's of AU \citep{Mason:2009}.

While the mass ratio distribution of the low mass stars is gaussian ($\mu \sim 0.6, \sigma \sim 0.1$), for more massive stars it is better represented by a power law $f(q) \propto q^{-0.4}$, where $q=m_1/m_2$ \citep{Kobulnicky:2007}. In both cases there is a bias towards more equal mass binaries at small separations.

In contrast, recent work on the young cluster Taurus-Auriga paints a different picture of the distribution \citep{Kraus:2011}. These authors find a much higher binarity fraction  of 73\% even among low mass stars ($< 2.5 M_\odot$), compared to 40-50\% in the field. While the separation distribution is consistent with a log-normal, as in the field, they find that lower mass systems ($M<0.7M_\odot$) have a tighter separation distribution than their higher mass counterparts. Other young clusters show an enhanced binary fraction for low mass stars \citep{Kraus:2008}, though Taurus is somewhat more extreme. 

\subsection{Empirical Models for Turbulent Fragmentation}
 Semi-empirical models such as \cite{Fisher04} have had some degree of success producing observed period and mass ratio distributions of low mass stars from the turbulent fragmentation mechanism. This work assumes that the components of binaries are drawn randomly from the initial mass function (IMF), and that there exists a constant efficiency factor with which mass and angular momentum are transferred from the core to the binary. They also draw eccentricities randomly from a thermal distribution, consistent with observations, in order to specify the orbit for a given binary angular momentum. A limitation of this work is that the physical process that governs the distribution of mass and angular momentum between binary components is ad hoc, and tuned to reproduce the observations.  While numerical work has shown that a multiple system can form out of a turbulent core, it remains to be shown in detail that the efficiency of angular momentum and mass removal scales as this model suggests. Moreover, while the final distribution of binary systems is consistent with the observation distribution of low mass stars,  it is not consistent with the orbital parameters and mass ratios for substellar companions and more massive stars. Recent AO surveys by \cite{Metchev:2009} show that the distribution of substellar companions to solar-type stars is inconsistent with being randomly drawn from the IMF. There is also a dearth of substellar mass companions at large distances.  On the opposite end of the mass spectrum there may also be inconsistencies. First, there is no explanation for the increase in binarity with core mass. Secondly, recent numerical work \citep{Krumholz2007a} has shown that radiation from the first star can suppress turbulent fragmentation in massive cores because more massive protostars are luminous even at early times.
 
 \subsection{Competitive Accretion and the IMF}
 Similarly, the competitive accretion picture has had success explaining portions of the observed initial mass function and some binary statistics  \citep{Moeckel:2010}. However, these models tend to overproduce the binaries containing the lowest mass stars and brown dwarfs, and have higher star formation efficiencies than observed \citep{Price:2009}. Another criticism is that the relative velocities between accreting cores and background gas are so high that the accretion rate onto the cores should be far too low to produce more massive stars \citep{Krumholz:2006}. In addition, the inter-core velocities appear to be much higher in the simulations than observed between actual star forming cores, calling into question whether or not there can be significant core-core interaction on the timescale of star formation \citep{Kirk:2010,Evans09}.
 
 The \cite{Goodwin:2005} picture does reproduce the observations for stars $\leq1 M _\odot$, but like the model of \cite{Fisher04} it has been tuned to do so. The mechanism which breaks some cores into two versus three stars, and the division of mass and angular momentum between them, is not specified in detail.

 \subsection{Disk Fragmentation: a possible solution to the luminosity and angular momentum problems}
Simulations now suggest that the disk scenario can produce a range of stellar separations and mass ratios, depending on primary mass, and the epoch of disk instability.  Although these models do not make clear predictions for the final separation and mass ratio distribution of stars, the disk fragmentation scenario may be consistent with two observational puzzles in star formation: the so-called luminosity problem \citep{Kenyon1990} and the angular momentum problem (see \citealt{Bodenheimer:1995} and references therein).

 \subsubsection{The Luminosity Problem}
The luminosity problem refers to the discrepancy between the observed and predicted bolometric luminosity of young protostars based on their accretion luminosities. Given a star formation timescale and stellar mass, we can predict the time-averaged mass accretion rate, and thus the corresponding luminosity. While uncertainties in the opacities and geometry of high extinction cores can complicate the interpretations, recent data from the Spitzer c2d survey confirms that most young sources are under-luminous compared to the expected luminosity from the time-averaged rate \citep{Evans09}. These authors suggest that variable, high accretion rates, particularly in the Class 0/I phase, are all but required by the data. Because so few cores are observed to have luminosities consistent with the time averaged rate (a few $10^{-6}M_\odot/$yr) accretion must occur in short bursts ($\sim 10^4$ years) well above this rate. However, a recent reanalysis has shown that such extreme variability may not be required. Instead, \cite{Offner:2011} suggest that a time variable accretion rate can explain the bulk of the variability.

For more massive protostars, high, variable accretion rates at early times are likely to produce fragmentation. However, such high mass cores are not typical among the sample described above, and thus the statistics on their observed accretion rates are limited \citep{Klaassen:2011}. Highly variable accretion from turbulent cores could further promote fragmentation if the disk is fed asymmetrically, which may also be consistent with observations (see e.g. \citealt{Tobin:2010}, and also \citealt{Stamatellos:2011}). 

\subsubsection{The Angular Momentum Problem}
The angular momentum problem refers to the fact that we do not understand how four orders of magnitude of angular momentum are repartitioned between the largest scales in Giant Molecular Clouds (GMCs) and the scales of binary orbits. 
 By the time dense cores have formed, line-width size observations reveal that the typical specific angular momentum has dropped by two orders of magnitude \citep{Goodman:1993}. When we look at binary stars, whose formation is intimately tied to the core angular momentum, the average specific angular momentum can drop another two orders of magnitude.  

Magnetic fields may play an important role in removing angular momentum on the largest scales \citep{Li:2004}, although hydrodynamic turbulence can be significant as well \citep{Jappsen:2004}. Disk fragmentation may be important on smaller scales: because disk size is regulated by the gas reservoir's angular momentum, if the disk fragments to form a binary or multiple before all of the infalling gas and angular momentum has accreted onto 
the star-disk system, binaries can form with less orbital angular momentum than that contained in the core. 
%Subsequent few-body interactions in disk-born higher order multiples can also redistribute angular momentum by expelling the lowest mass companions.
 
 \subsubsection{Other Observational Support for Disk Fragmentation}
Several other trends seem roughly consistent with the disk scenario. Both disk fragmentation and binary frequency increase with primary mass. At lower masses where disks are stabilized by stellar irradiation \citep{ML2005}, the turbulent cores out of which they form may be more susceptible to early fragmentation \citep{OKMKK10}  

The trend towards equal mass components at very high masses might be related to the fact that massive disks become unstable early on in their accretion histories \citep{PS2006}. There are also hints in the high mass multiplicity statistics that mass ratio scales inversely with separation \citep{Mason:2009}. Although migration can alter the orbits, disk fragmentation suggests a possible correlation between secondary mass and separation: fragmentation typically occurs towards the outer edge of the disk which is larger at late times, when there is also less mass available for the secondary to accrete.
 
 Another piece of evidence favorable for the disk scenario is the observation that binaries separated by less than roughly 40 AU are more likely to have the stellar spins aligned with the binary orbit \citep{Hale94}. Although this study does not account for higher order multiples, and has significant uncertainties in the measurements (e.g. sin i), this is suggestive that some objects were born in disks. Of course objects born from clouds with the same net rotation might also tend to be aligned. Recent observations of Herbig Ae/Be binaries also find alignment between the orbital plane of the binary and the accretion disks \citep{Wheelwright:2011}.
 
\section{Conclusions}
Despite their ubiquity, we still lack a coherent theory of binary and multiple star formation that can explain the enormous range of observed systems. This review provides a glimpse of the theoretical issues that must be resolved in order to make progress in the field.  Interested readers should find ample references to explore these topics further.

From a theoretical perspective, it remains difficult to predict the primordial distributions in mass, separation, and eccentricity, of binaries and multiples. These distributions must then be convolved with the known evolutionary processes in order to provide reliable initial conditions for binary stellar evolution calculations. At present, the observational data provides the best indication of the likely initial conditions for these models, despite lingering uncertainties regarding observational biases.

Recent work shows that disk fragmentation requires careful consideration as one of the viable mechanisms, especially for stars more massive than our sun. In particular, the combination of turbulent fragmentation across the mass spectrum with disk fragmentation added to the distribution at higher masses,  might explain the increase in multiplicity fraction with stellar mass.% To confirm this explanation, we must investigate whether or not the fraction of cores that produce sufficiently high disk masses for fragmentation are consistent with binary fractions as a function of primary mass. 

\acknowledgments The author would like to thank the conference organizers, and especially Linda Schmidtobreick for putting together an excellent and productive conference.

%\bibliography{diskbib_autohome}
\end{document}